\documentclass[twocolumn,noshowpacs,amsmath,amssymb]{revtex4}
\usepackage{graphicx}
\usepackage{amsfonts}
\usepackage{dcolumn}
\usepackage{bm}
\usepackage{color}
\usepackage{CJK}

\begin{document}
	
\newcommand{\gin}[1]{{\bf\color{blue}#1}}
\def\bc{\begin{center}}
\def\ec{\end{center}}
\def\bea{\begin{eqnarray}}
\def\eea{\end{eqnarray}}
\newcommand{\avg}[1]{\langle{#1}\rangle}
\newcommand{\Avg}[1]{\left\langle{#1}\right\rangle}

\title{Non-Markovian Majority-Vote model}
	
\author{Hanshuang Chen$^{1}$}\email{chenhshf@ahu.edu.cn}
	
\author{Shuang Wang$^{1}$}
	
\author{Chuansheng Shen$^{2}$}\email{csshen@mail.ustc.edu.cn}
	
\author{Haifeng Zhang$^{3}$}
	
\author{Ginestra Bianconi$^{4,5}$}
	
\affiliation{$^{1}$School of Physics and Materials Science, Anhui
		University, Hefei, 230601, China \\ $^{2}$School of Mathematics and
		Physics, Anqing Normal University, Anqing, 246133, China  \\
		$^{3}$School of Mathematical Science, Anhui University, Hefei,
		230601, China \\$^{4}$School of Mathematical Sciences, Queen Mary University of London,
		E1 4NS London, United Kingdom\\
		$^5$The Alan Turing Institute, The British Library, London, United Kingdom}

\date{\today}
	
\begin{abstract}	
Non-Markovian dynamics pervades human activity and social networks
and it induces memory effects and burstiness in a wide range of
processes including inter-event time distributions, duration of
interactions in temporal networks and  human mobility. Here we
propose a non-Markovian Majority-Vote model (NMMV) that introduces
non-Markovian effects in the standard (Markovian) Majority-Vote
model (SMV). The SMV model is one of the simplest two-state
stochastic models for studying opinion dynamics, and  displays a
continuous order-disorder phase transition at a critical noise. In
the NMMV model we assume that the probability that an agent changes
state is  not only dependent on the majority state of his neighbors
but it also depends on his {\em age}, i.e.\ how long the agent
has been in his current state. The NMMV model has two regimes: the
aging regime implies that the probability that an agent changes
state is decreasing with his age, while in the antiaging regime the
probability that an agent changes state is increasing with his age.
Interestingly, we find that the critical noise at which we observe
the order-disorder phase transition is a non-monotonic function of
the rate $\beta$ of the aging (antiaging) process. In particular
the critical noise in the aging regime displays a maximum as a
function of $\beta$ while in the antiaging regime displays a
minimum. This implies that the aging/antiaging dynamics can
retard/anticipate the transition and that there is an optimal rate
$\beta$ for maximally perturbing the value of the critical noise.
The analytical results obtained in the framework of the
heterogeneous mean-field approach are validated by extensive
numerical simulations on a large variety of network topologies.
\end{abstract}
	\maketitle
	
\section{Introduction}\label{sec1}
Many natural, social and technological  phenomena can be well
described by stochastic binary-state models formed by a large number
of interacting agents. Depending on the application, various types
of dynamical rules determining the stochastic switch of  the states
of the agents can be considered. This framework includes very well
known processes, such as the Ising model, the voter model and the
susceptible-infected-susceptible model, that have been used to model
magnetic materials  \cite{Baxter1989}, opinion formation
\cite{RMP09000591,SidneyRedner2017}, and epidemic spreading
\cite{RevModPhys.87.925,wang2016statistical},  among others
\cite{RevModPhys.90.031001,PR.687.1}. Strikingly, extensions or
modifications for the models can lead in a variety of cases to dynamical  behaviors
drastically different from the original ones. For example, the
presence of non-trivial structure in the interacting patterns such as
heavy-tailed degree distribution
\cite{RevModPhys.87.925,RMP08001275}, mesoscopic structures
\cite{PhysRep10000075,Chen2018JSTAT}, multilayer structures
\cite{boccaletti2014structure,bianconi2018multilayer,halu2013connect}, can induce significant change in the dynamics.
Moreover, relevant effect can be obtained also changing the dynamical
rules by introducing of more than two states
\cite{RevModPhys.54.235,Starnini.JSTAT2012}, time delay
\cite{PhysRevLett.118.168302}, nonhomogeneous interevent intervals
\cite{PhysRevE.84.015103,PhysRevE.84.036115,PhysRevE.83.036102,Min.EPL2013}, a fraction of zealot \cite{PhysRevLett.91.028701,PhysRevE.97.012310} or latency period \cite{PhysRevE.79.046107}.

The Majority-Vote (MV) model is a simple non-equilibrium Ising-like
system with up-down symmetry that presents an order-disorder phase
transition at a critical value of noise \cite{JSP1992}. The MV model
is also one of the paradigmatic models for studying opinion dynamics,
and it has been extensively studied in  regular lattices
\cite{PhysRevE.75.061110,PhysRevE.81.011133,PhysRevE.89.052109,PhysRevE.86.041123,PhysRevE.95.012101},
random graphs
\cite{PhysRevE.71.016123,PA2008}, and in complex networks including small-world networks
\cite{PhysRevE.67.026104,IJMPC2007,PA2015}, scale-free networks
\cite{IJMPC2006(1),IJMPC2006(2),PhysRevE.91.022816,Huang.EPL.2017}, modular
networks  \cite{CPL2015}, complete graphs  \cite{PhysRevE.96.012304},
and spatial networks  \cite{PhysRevE.93.052101}. Some
extensions were also proposed, such as multi-state MV models
\cite{PhysRevE.60.3666,JPA2002,PhysRevE.97.062304,JSM2010,JSM2016.073403,PA2012,PhysRevE.71.056124},
inertial effect
\cite{PhysRevE.95.042304,Chaos27.081102,PhysRevE.96.042305},
frustration due to anticonformists  \cite{Krawiecki2018EPJB}, and cooperation in
multilayer structures  \cite{Choi.NJP.2018,Liu.NJP.2019}.

Most of stochastic binary-state models are based on a memoryless
Markovian assumption, which implies that the switching rates from
one state to the other depend only on the present state of the
system. One of important properties of Markovian processes is that
the inter-event time intervals follow an exponential distribution
and the number of events  in a given time interval follows a
Poissonian distribution. The Markovian assumption facilitates
theoretical analysis of models. However, there is growing evidence
that human activity follows a non-Markovian dynamics driven by
memory effects. Non-Markovian bursty dynamics characterized by heavy
tail inter-event time distributions is ubiquitous in human
activities
\cite{Karsai2018,Barabasi.Nature2005,PhysRevE.73.036127,Barabasi.Nature2008,PhysRevLett.98.158702,PhysRevLett.103.038702,karsai2011small},and strongly affects the duration of interactions in temporal networks \cite{zhao2011entropy,PhysRevE.83.056109,PhysRevE.81.035101}.
Memory effects have also shown to be essential to model human
mobility and random walks over complex networks
\cite{Barabasi.Nature2008,dollarbill,rosvall2014memory}. Therefore, the
Markovian assumption provides only an approximate picture of the
real world.
	
In recent years, there is an increasing interest in understanding
the role of non-Markovian effects in stochastic binary-state models,
from the theoretical
\cite{PhysRevLett.110.108701,PhysRevX.4.011041,PhysRevLett.115.078701,PhysRevLett.118.128301,Kiss2015,Feng.NatCommu2019}
and from the numerical \cite{PhysRevE.90.042108,Masuda.SIAMRev2018}
perspective as well.

One important development of non-Markovian effects in stochastic binary-state models have been introduced by assuming that the switching probability between states depends on the age of the agent, i.e., how long an agent has been in its current state
\cite{PhysRevLett.101.018701,Eguiluz.SR2016}. The induced effects of
this non-Markovian dynamics are also called aging effects when the
switching probability decreases with the agent's age and antiaging
effects when the  switching probability increases with the agent's
age. These non-Markovian effects usually induce a slow-down of the relaxation dynamics toward the stationary state. In particular in social systems they can be related to behavioral inertia accounting for a tendency for a belief or an opinion to endure once formed.

A very important class of models describing opinion dynamics is the voter model and its variations. In the standard voter model, each agent updates his state by copying the state of one of his neighbors. The model exhibits ordering dynamics toward either of consensus states in finite-size systems \cite{SidneyRedner2017}. The effects of introducing a non-Markovian dynamics within the voter model and its variations have been considered in several works.
In Ref.~\cite{PhysRevLett.101.018701}, Stark \emph{et al.}
reported a counter-intuitive phenomenon induced by aging in the voter model. They showed that the transition probability between two
opposite states decreases with age, but the time to reach a
macroscopically ordered state can be accelerated. In Ref.~\cite{Toral.PA2020}, Peralta \emph{et al.} studied systematically
the aging version of the voter model at the mean-field level, and they
showed that the model reaches consensus or gets trapped in a frozen
state depending on the specific form describing the transition probability
and the nodes' age. They also considered the antiaging case when the
transition probability is an increasing function of age. For the
latter case, the model always reaches a steady state with
coexistence of two states. In the noisy voter model, additional stochastic effects are introduced in the opinion dynamics. In particular given an agent of a noisy voter model and his randomly selected neighbor, the agent does not adopt the neighbor opinion deterministically. An important consequence of this is that  a stationary state can be achieved without consensus  \cite{Peralta2018}.
In Refs.~\cite{PhysRevE.98.032104,Artime2019}, it has been shown that the
aging effects in the noisy voter model can alter the  character of the
phase transition. In the absence of aging, the model show a
finite-size discontinuous transition between ordered and disordered
phases. When the aging is introduced, the transition becomes a well defined second order transition observed in the thermodynamic limit. Moreover,
recently Peralta \emph{et al.} in Ref.~\cite{Toral.JSTAT2020} proved
that the non-Markovian noisy voter model can be approximately
reduced to a non-linear noisy voter model which is Markovian.

In the present work, we reveal the role of non-Markovian dynamics in
the Majority-Vote (MV) model providing results that enrich the scenario depicted by the works above summarized. In the MV model each agent tends to agree with the majority state of his neighbors, and disagreement only occurs with  probability $f$. Here $f$ can be interpreted as the internal noise due to imperfect information exchange or uncertainty on the states of neighbors.
As $f$ increases, the MV model shows a continuous order-disorder transition belonging to the universality class of the equilibrium Ising model \cite{JSP1992}. In particular, for $f=0$ the MV model is equivalent to the zero-temperature Ising model with Glauber dynamics \cite{PNAS2005}.
		
It is interesting to discuss the difference between the MV model and the voter model and its variations. The main difference of the MV model with respect to the  voter model is that at each time in the MV model each agent changes opinion depending on the {\em majority} of its neighbors while in the voter model each agent changes opinion depending on the state of a {\em single randomly selected neighbor}. Moreover in the standard voter model the system reaches consensus while this is not the case in the MV model.
The noisy voter model is closer to  MV model as in both models we can reach a stationary state with a majority opinion but without consensus. However due to the different dynamical rules the nature of the phase transition observed in the two models is different as demonstrated by the  different universality class of the ordering dynamics of the voter model \cite{PhysRevLett.94.178701,PhysRevE.71.066107}. 
	
Here we propose the non-Markovian Majority-Vote (NMMV) model by incorporating non-Markovian dynamics in
the Majority-Vote model. In the NMMV model, the transition
probability between states not only depends on the majority state of
the agent’s neighbors and noise inetensity $f$, but also depends on the agent's age. Specifically, the NMMV model includes two regimes: the aging regime in which the
probability of a state switch decreases with the agent's age and an
antiaging regime in which the probability of a state switch
increases with the agent's age. We indicate with $\beta$ the rate of
change of the transition probability with age. The NMMV model also displays a continuous phase transition as a function of $f$: for
$f<f_c^{NMMV}$ the NMMV model is in the ordered phase, i.e.,
the network displays a clear majority state, for $f\geq f_c^{NMMV}$
the NMMV is in the disordered phase where no global majority state exist. We show that the non-Markovian
dynamics strongly affects the value of the critical noise
$f_c^{NMMV}$. In particular, in the aging regime the non-Markovian
dynamics retards the transition with respect to the standard
Majority-Vote model (SMV) and the critical noise $f_c^{NMMV}$ in the
NMMV model is larger or equal to the critical noise $f_c^{SMV}$ in
the SMV model, i.e., $f_c^{NMMV}\geq f_c^{SMV}$. In the
antiaging regime, instead, the relation between the critical noise
in the NMMV model and in the SMV model are reversed, i.e.,
$f_c^{NMMV}\leq f_c^{SMV}$. Interestingly, by solving the model in
the framework of an heterogeneous mean-field approach, we can derive
analytically the non-monotonic dependence of  the critical noise
$f_c^{NMMV}$ on the rate $\beta$.  In the aging regime, the critical
noise displays a maximum  at a non-zero but finite value of $\beta$.
In the antiaging regime, a minimum of the critical noise as a
function of $\beta$ is found. This means that the non-Markovian
dynamics can be used to retard or anticipate the transition.
	
The theoretical mean-field predictions are in good agreement with extensive simulations of the model.
	
The paper is structured as follows: in Sec.~\ref{sec2} we define the
NMMV model; in Sec.~\ref{sec3} we present the analytic solution of
the model obtained in the framework of the heterogeneous mean-field
approach; in Sec.~\ref{sec4} we characterize the critical properties
of the model including the analytical expression of the critical
noise, and its dependence on the rate $\beta$; in Sec.~\ref{sec5} we
compare the analytic predictions to the simulation results; finally
in Sec.~\ref{sec6} we provide the conclusions.
	
\section{Majority-Vote Model with non-Markovian switching of states}\label{sec2}

In this section we introduce the non-Markovian Majority-Vote model which differs from the standard Majority-Vote model
\cite{JSP1992} by introducing a non-Markovian mechanism for the
switching of states. Therefore in the NMMV model the agents have a
probability of switching states that depends on their {\em age},
i.e.\ for how long they have been in their current state.
	
We consider a population of $N$ agents defined on a static network
topology. Each agent $i$ with $i=1,\cdots,N$ is located on a node
$i$ of the network. Each agent  is assigned two dynamical variables:
a binary variable $\sigma_i=\pm 1$ ({\em his state}) describing the
agent's opinion/vote and a variable $a_i$ ({\em his age})
indicating for how long the agent has not changed his state.
Initially the states $\{\sigma_i\}$ are randomly assigned to the
agents and the variables $\{a_i\}$ are initialized by setting
$a_i=0$ for every agent $i$ of the network. At each time step, an
agent $i$ is chosen at random and his state is switched with
probability  $w_i$ which implements the non-Markovian Majority Vote
process. Thus with  probability $w_i$, the agent $i$ switches state
and  the age of agent $i$ is reset to zero, i.e., \bea
\sigma_i &\rightarrow&-\sigma_i,\nonumber \\
a_i&\rightarrow & 0.
\eea
Otherwise, nothing happens except for the age increased by one, i.e.,
\bea
a_i\rightarrow a_i+1.
\eea
In both cases the time is updated  according to
\bea
t\to t+\Delta t,
\eea
with $\Delta t=1/N$.
The richness of the model resides on the definition of the switching probability $w_i$ given by
\begin{eqnarray}
w_i = \nu_i w_i^{SMV}, \label{eq2}
\end{eqnarray}
where $0\leq \nu_i \leq 1$, called the {\em activation probability},
is a function of  the age $a_i$  of agent $i$ and where $w_i^{SMV}$
is the switching probability in the SMV model, i.e., it is
independent of the age variable. The  contribution  $w_i^{SMV}$ to
the switching probability $w_i$ of the agent $i$ depends on the
majority state of $i'$s  neighborhood and on a parameter $f$
called the {\em noise intensity}. If the state $\sigma_i$ of the
agent is opposite to the majority state of his neighbors, $w_i^{SMV}$
contributes to the switching probability to the majority state by a
term $1-f$. If the state $\sigma_i$ of the agent is the same as  the
majority state of his neighbors, $w_i^{SMV}$ contributes to the
switching probability to the majority state by a term $f$. If there
is no clear majority of the agent $i$'s neighbors, i.e., half of
the neighbors have state $\sigma_j=+1$ and half of the neighbors
have state $\sigma_j=-1$, then $w_i^{SMV}=1/2.$ Therefore,
$w_i^{SMV}$ can be expressed as
\begin{eqnarray}
{w_i^{SMV}}{\rm{ = }}\frac{{\rm{1}}}{{\rm{2}}}\left[ {{\rm{1 -
}}\left( {{\rm{1 - 2}}f} \right){\sigma _i}S\left( {\sum\limits_{j
\in {\mathcal {N}_i}} {{\sigma _j}} } \right)} \right], \label{eq1}
\end{eqnarray}
where $\mathcal {N}_i$ denotes the set of neighbors of agent $i$,
and $S(x)$, defined as $S(x)=\mbox{sgn}(x)$ if $x\neq0$ and $S(0)=0$, indicates the majority state of his neighborhood.
	
The NMMV model reduces to the SMV model in the case in which we
consider a trivial choice of $\nu_i$, i.e.\ $\nu_i=1$ for all
agent $i$. In this case, as $f$ increases, the model undergoes a
continuous order-disorder phase transition at a critical value of
noise intensity $f=f_c^{SMV}$ \cite{PhysRevE.91.022816}.
	
However, in a number of real scenarios for social and human dynamics  it has been shown that non-Markovian effects are relevant
\cite{Karsai2018}. Indeed a large number of human activity including written correspondence, emails \citep{Barabasi.Nature2005}, mobile phone communication \cite{zhao2011entropy}  is not memoryless, on the contrary it is characterized by important non-Markovian effects typically leading to intermittent and bursty dynamics.
		
Different models for explaining the emergence of bursty dynamics have been proposed (see for a review Ref.\cite{Karsai2018}).
Interestingly, a  model \cite{PhysRevE.81.035101,PhysRevE.83.056109,zhao2011entropy} explaining the occurrence of  bursty human dynamics of social interactions assumes that  a number of feedback mechanisms affect human behavior, which  introduce memory effects in the rate at which  an agent to change his state. In particular in this framework it is assumed that  each agent does not change his state at a constant rate in time, rather the rate at which he changes his state depends on  the time elapsed since he adopted his current state. This framework, originally proposed to model the duration of social interactions is a very general framework that can be also applied to opinion dynamics. In opinion dynamics this framework will give rise to  a simple  yet very general phenomenological model to describe the inertia of the agents in retaining their own opinion.
By following these considerations, here we  capture the effect of the  non-Markovian opinion dynamics  in the MV model by  assuming that the probability $\nu_i$ at which  an agent $i$ changes opinion depends on how long the agent has retained its current opinion, i.e.
\bea
\nu_i=\nu(a_i),
\eea
where $a_i$ indicates the {\em age} of agent $i$.
In the following  we will consider several different functional forms for the function $\nu(a)$ including  exponential, linear, rational, and power-law dependence with the the age $a$. To start with a  concrete example let us now consider the  exponential form for
$\nu(a)$, given by
\begin{eqnarray}
		\nu(a)  = \left( {{\nu _0} - {\nu _\infty }} \right){e^{ - \beta a }} + {\nu _\infty },\label{eq3}
\end{eqnarray}
where $\nu_0=\nu(0)$ and $\nu_\infty=\lim_{a\to \infty}\nu(a)$.
		
The probability $\nu(a)$ capture the non-Markovian nature of the dynamics and is parametrized by the parameter $\beta>0$. Note that $\beta$ characterizes the rate of exponential change of $\nu$ as a function of  $a$. Obviously, in the limits of
$\beta \rightarrow 0$ and $\beta \rightarrow \infty$, all the agents have the same fixed value of activity,
$\nu \equiv \nu_0$ and $\nu \equiv \nu_\infty$, and the dynamics is thus equivalent to the SMV model with the time scaled by a factor $\nu_0^{-1}$ and $\nu_\infty^{-1}$,
respectively.
	
We distinguish two different regimes of the dynamics:
\begin{itemize}
		\item[(i)] {\em Aging regime}.  For $\nu_0>\nu_\infty$, $\nu(a)$ decays exponentially with $a$, implying that the longer an agent is in a given state, the more difficult is for him to change state.

		\item[(ii)] {\em Antiaging regime}. For  $\nu_0<\nu_\infty$,  $\nu(a)$ increases
		exponentially with $a$. Such an
		case can be interpreted as ``rejuvenating" dynamics where  agents  become more prone to change state as they are longer on a given state.
\end{itemize}
Without loss of generality, set equal to one the maximum between $\nu_0$ and $\nu_{\infty}$, i.e.\ we put $\max \left\{ {{\nu _0},{\nu _\infty }}
\right\} = 1$. Moreover, to avoid trivial frozen states of the dynamics, the minimum between  $\nu_0$ and $\nu_\infty$ is set to be larger than zero,
i.e., $\min \left\{ {{\nu _0},{\nu _\infty }} \right\}	>0$.

\section{Heterogeneous mean-field solution of the model}\label{sec3}
	
In order to capture the phase diagram of the NMMV model on a random network with given degree distribution $P(k)$, we solve the model using the heterogeneous mean-field approach ~\cite{RMP08001275}.
Therefore we assume that the probability that an agent $i$ is in a given state depends exclusively on his degree $k$ and his age $a$ and we denote by $x_{k,a}^\pm$ the probability that an agent of degree $k$ has age $a$ and is in the state $\pm 1$.
It follows that the probability $x_k^ \pm$ of an agent of degree $k$ in the state $\pm 1$,
is given by
\bea
x_k^ \pm  = \sum\nolimits_{a = 0}^\infty  {x_{k,a}^ \pm}.
\eea

In order to solve the dynamical equations of the NMMV model in the heterogeneous mean-field approximation we also need to evaluate the switching probability $w_{k,a}^{\pm}$ of an agent of degree $k$ and age $a$.
Let us define  ${{\tilde x}^ \pm }$  the probability that by following a link we reach a node in state $\pm 1$,  given by
\begin{eqnarray}
	{{\tilde x}^ \pm } = \sum\limits_k {\frac{{kP\left( k
				\right)}}{{\left\langle k \right\rangle }}x_k^ \pm }  =
	\sum\limits_k {\frac{{kP\left( k \right)}}{{\left\langle k
				\right\rangle }}\sum\limits_{a = 0}^\infty  {x_{k,a}^ \pm } }.\label{eq4}
\end{eqnarray}
For a node of degree $k$, the probability that the majority state
among his neighborhoods is $\pm 1$ is given by the binomial distribution,
\begin{eqnarray}
	{\psi _k}\left( {{{\tilde x}^ \pm }} \right) = \sum\limits_{n =
		\left\lceil {k/2} \right\rceil }^k {\left( {1 - \frac{1}{2}{\delta
				_{n,k/2}}} \right)C_k^n} {\left( {{{\tilde x}^ \pm }}
		\right)^n}{\left( {1 - {{\tilde x}^ \pm }} \right)^{k - n}}, \nonumber \\ \label{eq5}
\end{eqnarray}
where $\left\lceil  \cdot  \right\rceil$ is the ceiling function, $\delta_{r,s}$ is the Kronecker symbol, and
$C_k^n = {{k!} \mathord{\left/
{\vphantom {{k!} {\left[ {n!\left( {k - n} \right)!} \right]}}} \right.
\kern-\nulldelimiterspace} {\left[ {n!\left( {k - n} \right)!} \right]}}$
are the binomial coefficients.
According to Eq.(\ref{eq2}), we can write down the switching probability $w_{k,a}^\pm$ of
an agent of state $\pm 1$ with degree $k$ and age $a$ as
\begin{eqnarray}
w_{k,a}^ \pm  = \nu \left( {a + 1} \right){\Psi _k}\left( {{{\tilde
				x}^ \pm }} \right),\label{eq6}
\end{eqnarray}
where $\nu(a+1)$ is given by Eq.(\ref{eq3}), and ${\Psi _k}\left(
{{{\tilde x}^ \pm }} \right)$ is the flipping probability of an
agent of state $\pm 1$ without the aging effect	~\cite{PhysRevE.91.022816}, i.e.,
\begin{eqnarray}
	{\Psi _k}\left( {{{\tilde x}^ \pm }} \right) = \left( {1 - f}
	\right)\left[ {1 - {\psi _k}\left( {{{\tilde x}^ \pm }} \right)}
	\right] + f{\psi _k}\left( {{{\tilde x}^ \pm }} \right).\label{eq7}
\end{eqnarray}
	
The dynamical equations that determine the time evolution of the
probabilities $x_{k,a}^\pm$ are a function of the switching
probabilities $w_{k,a}^\pm$. These equations can be deduced by
observing that at each time step one of the following four possible
events occurs.
	
\begin{itemize}
\item[(i)] An agent in state $+1$ having degree $k$ and age $a$ is chosen and his  state is flipped. The rate at which $x_{k,a}^{+}$ decreases and
$x_{k,0}^{-}$ increases due to this process is $x_{k,a}^{+} w_{k,a}^{+}$.

\item[(ii)] An agent in state $+1$ having degree $k$ and age $a$ is chosen
but his state is not flipped. The rate at which $x_{k,a}^{+}$ decreases and
$x_{k,a+1}^{+}$ increases  due to this process is $x_{k,a}^{+} (1-w_{k,a}^ {+})$.

\item[(iii)] An agent in state $-1$ having degree $k$ and age $a$ is chosen
and the state is flipped. The rate at which $x_{k,a}^{-}$ decreases and
$x_{k,0}^{+}$ increases  due to this process is $x_{k,a}^{-} w_{k,a}^ {-}$.

\item[(iv)] An agent in state $-1$ having degree $k$ and age $a$ is chosen
but the state is not flipped. The rate at which $x_{k,a}^{-}$ decreases and
$x_{k,a+1}^{-}$ increases  due to this process is  $x_{k,a}^{-} (1-w_{k,a}^ {-})$.
\end{itemize}
	
Accordingly, the rate equations for $x_{k,a}^\pm$ read
\begin{eqnarray}
	\frac{{dx_{k,0}^ + }}{{dt}} &= &\sum\limits_{a = 0}^\infty  {x_{k,a}^ - w_{k,a}^ - }  - x_{k,0}^ +, \label{eq8} \\
	\frac{{dx_{k,a}^ + }}{{dt}} &=& x_{k,a - 1}^ + \left( {1 - w_{k,a - 1}^ + } \right) - x_{k,a}^ + ,{\kern 10pt}  a \geq1, \label{eq9} \\
	\frac{{dx_{k,0}^ - }}{{dt}} &=& \sum\limits_{a = 0}^\infty  {x_{k,a}^ + w_{k,a}^ + }  - x_{k,0}^ -, \label{eq10}\\
	\frac{{dx_{k,a}^ - }}{{dt}} &=& x_{k,a - 1}^ - \left( {1 - w_{k,a - 1}^ - } \right) - x_{k,a}^ - ,{\kern 10pt}  a \geq1. \label{eq11}
\end{eqnarray}

In stationary state, by setting the time derivative of $x_{k,a}^\pm$ equal to
zero, we obtain that the probabilities $x_{k,a}^\pm$ obey
	
\begin{eqnarray}
	x_{k,0}^ +  &=& \sum\limits_{a = 0}^\infty  {x_{k,a}^ - w_{k,a}^ - },\label{eq12} \\
	x_{k,a}^ +  &=& x_{k,a - 1}^ + \left( {1 - w_{k,a - 1}^ + } \right),{\kern 10pt}  a \geq1, \label{eq13}  \\
	x_{k,0}^ -  &=& \sum\limits_{a = 0}^\infty  {x_{k,a}^ + w_{k,a}^ + }, \label{eq14}\\
	x_{k,a}^ -  &=& x_{k,a - 1}^ - \left( {1 - w_{k,a - 1}^ - } \right),{\kern 10pt} a\geq1.\label{eq15}
\end{eqnarray}
	
Using Eq.(\ref{eq13}) and Eq.(\ref{eq14}), and summing $x_{k,a}^{+}$
over the values of $a$ greater or equal to one we get
\begin{eqnarray}
	x_{k,0}^{+}=x_{k,0}^{-}.\label{eq16}
\end{eqnarray}
This condition is a necessary condition for stationarity. In fact, at  stationarity the probability that a node is in a given state does not change with time, or equivalently the expected number of
agents in state $+1$ that change their state  (and reset their age to $a=0$) should be equal to the number of agent in state  $-1$  that change their state  (and reset their age to $a=0$)   \cite{PhysRevE.98.032104}.
	
In terms of Eq.(\ref{eq13}) and Eq.(\ref{eq15}),
$x_{k,a}^\pm$ for $a\geq1$ can be computed in a recursive way, and
then are expressed by $x_{k,0}^\pm$,
\begin{eqnarray}
\begin{array}{lr}
	x_{k,a}^ \pm  = x_{k,0}^ \pm  {F_{k,a}}\left( {{{\tilde x}^ \pm }}
	\right), & a\geq1,\label{eq17}
\end{array}
\end{eqnarray}
where for convenience we have introduced the function $F_{k,a}$, given by
\begin{eqnarray}
	{F_{k,a}}\left( {{{\tilde x}^ \pm }} \right) = \prod\limits_{j =
		0}^{a - 1} {\left[ {1 - w_{k,j}^ \pm \left( {{{\tilde x}^ \pm }}
			\right)} \right]}.\label{eq18}
\end{eqnarray}
	
Substituting Eq.(\ref{eq17}) into the definition $x_k^ \pm  = \sum\nolimits_{a = 0}^\infty  {x_{k,a}^ \pm }$, we have
\begin{eqnarray}
	x_k^ \pm  = x_{k,0}^ \pm F_k\left( {{{\tilde x}^ \pm }} \right),\label{eq19}
\end{eqnarray}
with
\begin{eqnarray}
	F_k\left( {{{\tilde x}^ \pm }} \right){\rm{ = 1 + }}\sum\limits_{a = 0}^\infty  {{F_{k,a}}\left( {{{\tilde x}^ \pm }} \right)} .\label{eq20}
\end{eqnarray}
In order to find  $x_k^+$ we note that by using Eq.(\ref{eq16}),  we can express the ratio $x_k^+/x_k^-$ as
\begin{eqnarray}
	\frac{{x_k^ + }}{{x_k^ - }} =  \frac{{x_{k,0}^ + }}{{x_{k,0}^ -
	}}\frac{{F_k\left( {{{\tilde x}^{\rm{ + }}}} \right)}}{{F_k\left(
			{{{\tilde x}^ - }} \right)}} = \frac{{F_k\left( {{{\tilde x}^{\rm{ +
			}}}} \right)}}{{F_k\left( {{{\tilde x}^ - }} \right)}}.\label{eq21}
\end{eqnarray}
Substituting ${\tilde x}^-$ with $1-{\tilde x}^+$ in Eq.(\ref{eq21}), we then obtain
\begin{eqnarray}
	x_k^ +  = \frac{{F_k\left( {{{\tilde x}^{\rm{ + }}}}
			\right)}}{{F_k\left( {{{\tilde x}^{\rm{ + }}}} \right) + F_k\left(
			{1 - {{\tilde x}^ + }} \right)}}.\label{eq22}
\end{eqnarray}
Finally by using  Eq.(\ref{eq22}) in the left-hand side of  Eq.(\ref{eq4}),  we find the self-consistent equation  of $\tilde x^+$,
\begin{eqnarray}
	{{\tilde x}^ + } = \sum\limits_k {\frac{{kP\left( k
				\right)}}{{\left\langle k \right\rangle }}\frac{{F_k\left( {{{\tilde
							x}^{\rm{ + }}}} \right)}}{{F_k\left( {{{\tilde x}^{\rm{ + }}}}
				\right) + F_k\left( {1 - {{\tilde x}^ + }} \right)}} }.\label{eq23}
\end{eqnarray}
This equation can be solved numerically by finding $\tilde x^+$ by
iterating Eq.(\ref{eq23}) starting from an initial value of $\tilde
x^+\neq 1/2$. Once $\tilde x^+$ is found, we can calculate $x_k^ +$
by using Eq.(\ref{eq22}). This allow us to find  the average
magnetization per node by \bea m = \sum\nolimits_k {P\left( k
\right)\left( {x_k^ +  - x_k^ - } \right)}  = \sum\nolimits_k
{P\left( k \right)\left( {2x_k^ +  - 1} \right)}. \eea
	
This theoretical treatment of the model provides predictions that
can be compared to simulation results revealing the critical
properties of the NMMV model. In particular, the main features of the
steady state configurations can be described by plotting $m$ as a
function of $f$ for different values of $\beta$.
	
\begin{figure}
\centerline{\includegraphics*[width=1.0\columnwidth]{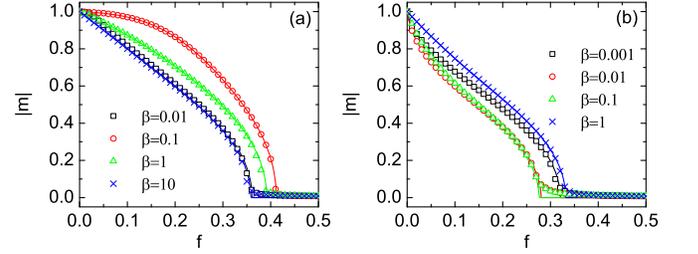}}
\caption{The absolute value of $m$, $|m|$, is plotted  as a function
			of the noise $f$ for several values of $\beta$. Panel (a)  shows
			$|m|$ versus $f$  for $\nu_0>\nu_\infty$, i.e.\ for a
			dynamics in the aging regime; panel (b) shows $|m|$ versus $f$ for
			$\nu_0<\nu_\infty$, i.e.\ for a dynamics in the antiaging
			regime. The simulations (symbols) performed on  a regular random
			network (RR) with $N=10^4$ nodes and with degree of the nodes given
			by $\left\langle k \right\rangle=20$ are compared with theoretical
			predictions (solid lines). All results are obtained for $\max
			\left\{ {{\nu _0},{\nu _\infty }} \right\} = 1$ and $\min \left\{
			{{\nu _0},{\nu _\infty }} \right\} = 0.05$.
\label{fig1}}
\end{figure}

In Fig.\ref{fig1}(a), we report such results for
$\nu_0>\nu_\infty$, when the non-Markovian dynamics is in the aging
regime. Here we have used regular random networks (RR) whose degree
distribution follows a delta function, $P(k) = \delta \left( {k -
\left\langle k \right\rangle } \right)$ with $\left\langle k
\right\rangle=20$ and network size $N=10^4$. Direct simulation
results are compared to theoretical predictions finding excellent
agreement (see Fig.\ref{fig1}). The order parameter $|m|$ shows a
continuous second-order phase transition as noise intensity $f$
varies, similar to the SMV model. The transition point, i.e.,
the critical value of noise intensity $f_c^{NMMV}$, depends on the value of $\beta$. In the aging regime, as $\beta$ increases, $f_c^{NMMV}$ displays a maximum at $\beta=\beta_m^{aging}$. In the antiaging regime ($\nu_0<\nu_\infty$), $f_c^{NMMV}$ shows again a non-monotonous behavior but instead of displaying a maximum as a function of $\beta$ (like in presence of the aging dynamics) it displays a minimum at $\beta=\beta_m^{antiaging}$ (see Fig.\ref{fig1}(b)).

\section{The phase diagram}\label{sec4}
\subsection{The critical noise}\label{sec4.1}
In this paragraph we will use the heterogeneous mean-field approach
to  derive the expression for the critical noise $f_c^{NMMV}$ in the NMMV model. First of
all, we notice that $\tilde x^+=1/2$,  is always a solution of
Eq.(\ref{eq23}). This state corresponds to the disordered phase
where the state of each agent is totally random. Such a trivial
solution loses its stability when the noise intensity is less than a
critical value, i.e., $f<f_c^{NMMV}$. According to linear stability
analysis, the critical noise $f_c^{NMMV}$ can be found by imposing that the derivative of the right-hand side of Eq.(\ref{eq23}) with respect to $\tilde
x^+$ calculated for $\tilde x^+=1/2$ is equal to one, i.e.,
$f_c^{NMMV}$ satisfies
\begin{eqnarray}
  \sum\limits_k {\frac{{kP\left( k \right)}}{{\left\langle k
				\right\rangle }}\frac{{F'_k\left( {\frac{1}{2}}
				\right)}}{{2F_k\left( {\frac{1}{2}} \right)}}}  = 1.\label{eq24}
\end{eqnarray}
At $\tilde x^+=1/2$, $\psi_k$ and also $\Psi_k$ are independent of $k$. In particular we have ${\Psi _k}\left( {\frac{1}{2}} \right) = \frac{1}{2}$ for all value of $k$.
Therefore using  Eq.(\ref{eq20}),  this implies that also $F_k\left( {\frac{1}{2}} \right)$ is independent of $k$ and is given by
\begin{eqnarray}
	{F}\left( {\frac{1}{2}} \right) = 1 + \sum\limits_{a = 1}^\infty  {{F_{a}}} \left( {\frac{1}{2}} \right),\label{eq25}
\end{eqnarray}
with
\begin{eqnarray}
	{F_{a}}\left( {\frac{1}{2}} \right) = \prod\limits_{j = 1}^a {\left(
		{1 - \frac{1}{2}\nu \left( j \right)} \right)},\label{eq26}
\end{eqnarray}
(note that here we have omitted the subscript $k$ in the expression
of $F_{k}(\frac{1}{2})$ and $F_{k,a}(\frac{1}{2})$ as they do not
depend on $k$.) After some simple algebra, we can express $F'_k\left( {\frac{1}{2}}
\right)$ as
\begin{eqnarray}
	F'_k\left( {\frac{1}{2}} \right) =  - \Psi_k '\left( {\frac{1}{2}}
	\right)\sum\limits_{a = 1}^\infty  {{F_{a}}\left( {\frac{1}{2}}
		\right)\sum\limits_{j = 1}^a {\frac{{{\nu(j)}}}{{1 - \frac{1}{2}{\nu
						(j)}}}} },\label{eq27}
\end{eqnarray}
with
\begin{eqnarray}
	\Psi '_k\left( {\frac{1}{2}} \right) = \left( {2f - 1} \right)\psi
	'_k\left( {\frac{1}{2}} \right),\label{eq28}
\end{eqnarray}
and
\begin{eqnarray}
	\psi '_k\left( {\frac{1}{2}} \right) = {2^{1 - k}}kC_{k -
		1}^{\left\lceil {\left( {k - 1} \right)/2} \right\rceil }.\label{eq29}
\end{eqnarray}
Substituting Eqs.(\ref{eq25}-\ref{eq29}) into Eq.(\ref{eq24}), we obtain the critical noise $f_c^{NMMV}$ in the NMMV model,
\begin{eqnarray}
	f_c^{NMMV} = \frac{1}{2} - G\left( {\beta ;{\nu _0},{\nu _\infty }}
	\right)\frac{{\left\langle k \right\rangle }}{{\sum\nolimits_k
			{{k^2}P\left( k \right){2^{1 - k}}C_{k - 1}^{\left\lceil {\left( {k
							- 1} \right)/2} \right\rceil }} }}, \nonumber \\ \label{eq30}
\end{eqnarray}
where
\begin{eqnarray}
	G\left( {\beta ;{\nu _0},{\nu _\infty }} \right) = \frac{{F\left(
			{\frac{1}{2}} \right)}}{{\sum\limits_{a = 1}^\infty  {{F_a}\left(
				{\frac{1}{2}} \right)\sum\limits_{j = 1}^\infty  {\frac{{{\nu(j)}}}{{1 - \frac{1}{2}{\nu(j)}}}} } }}.\label{eq31}
\end{eqnarray}
Using Stirling's approximation for large $k$,
	$C_{k -
		1}^{\left\lceil {\left( {k - 1} \right)/2} \right\rceil } \approx
	{{{2^{k - 1}}} \mathord{\left/ {\vphantom {{{2^{k - 1}}} {\sqrt
						{{{k\pi } \mathord{\left/ {\vphantom {{k\pi } 2}} \right.
									\kern-\nulldelimiterspace} 2}} }}} \right.
			\kern-\nulldelimiterspace} {\sqrt {{{k\pi } \mathord{\left/
						{\vphantom {{k\pi } 2}} \right.
						\kern-\nulldelimiterspace} 2}} }}$,
Eq.(\ref{eq30}) can be simplified to
\begin{eqnarray}
	f_c^{NMMV} = \frac{1}{2} - G\left( {\beta ;{\nu _0},{\nu _\infty }}
	\right)\sqrt {\frac{\pi }{2}} \frac{{\left\langle k \right\rangle
	}}{{\left\langle {{k^{3/2}}} \right\rangle }},\label{eq32}
\end{eqnarray}
where $\langle\ldots\rangle$ denotes the average over the degree distribution $P(k)$.
The critical noise $f_c^{NMMV}$ dependence on the non-Markovian dynamics is fully captured by the function $G\left( {\beta ;{\nu _0},{\nu _\infty }} \right)$, which can be considered as a function of $\beta$ for any given value of the parameters  $\nu_0$ and $\nu_{\infty}$.
We distinguish two main regimes:
\begin{itemize}
		\item[(i)] For $\nu_0>\nu_{\infty}$, $G\left( {\beta ;{\nu _0},{\nu _\infty }} \right)$ captures the dependence of $f_c^{NMMV}$ on $\beta$ in the  {\em aging regime};
		\item[(ii)]For $\nu_0<\nu_{\infty}$, $G\left( {\beta ;{\nu _0},{\nu _\infty }} \right)$ captures the dependence of $f_c^{NMMV}$ on $\beta$ in the {\em antiaging regime}.
\end{itemize}
When the aging effects are not
taken into account, $\nu(a)\equiv \nu$, $G\left( {\beta ;{\nu_0},{\nu _\infty }} \right) = \frac{1}{2}$, and Eq.(\ref{eq32})  thus
reduces to the expression of the critical noise in the SMV model \cite{PhysRevE.91.022816},
\begin{eqnarray}
	f_c^{SMV} = \frac{1}{2} - \frac{1}{2}\sqrt {\frac{\pi }{2}}
	\frac{{\left\langle k \right\rangle }}{{\left\langle {{k^{3/2}}}
			\right\rangle }}.\label{eq33}
\end{eqnarray}
	
\subsection{The function $G\left( {\beta ;{\nu _0},{\nu _\infty }}\right)$ }\label{sec4.2}

\begin{figure}
\centerline{\includegraphics*[width=1.0\columnwidth]{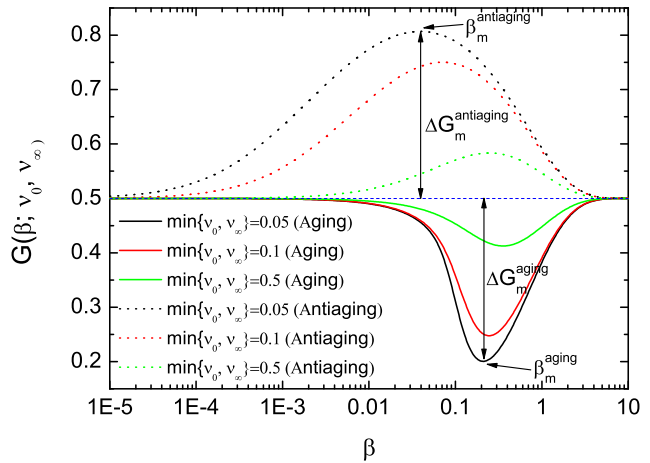}}
\caption{The function $G\left( {\beta ;{\nu _0},{\nu _\infty }} \right)$ defined in Eq.(\ref{eq31}) is plotted as a function of $\beta$ in  the aging regime
(for $\nu_0>\nu_{\infty}$) and in the antiaging regime (for $\nu_0<\nu_{\infty}$). All curves are obtained for $\max \left\{ {{\nu _0},{\nu _\infty }} \right\}= 1$.
\label{fig2}}
\end{figure}

As noted before, the function  $G\left( {\beta ;{\nu _0},{\nu _\infty }}
\right)$ captures all the  dependence of the critical noise $f_c^{NMMV}$ on the non-Markovian dynamics. In particular, from Eq.(\ref{eq32}) and Eq.(\ref{eq33}) we deduce that the function $G\left( {\beta ;{\nu _0},{\nu _\infty }}
\right)$ characterizes the relation between the critical noise in NMMV model and in the SMV model. In fact, we have
\bea
	2G\left( {\beta ;{\nu _0},{\nu _\infty }}\right) = \frac{1/2 - f_c^{NMMV}}{1/2 - f_c^{SMV}},
	\label{eq:G}
\eea
The numerical solution of Eq.(\ref{eq31}) reveals that the function
$G\left( {\beta ;{\nu _0},{\nu _\infty }} \right)$ displays a
non-monotonous behavior as a function of $\beta$ when $\nu_0$ and
$\nu_{\infty}$ are fixed to a constant value. In particular, the
function $G\left({\beta ;{\nu _0},{\nu _\infty }} \right)$ displays
a minimum as a function of $\beta$ in the aging regime and a maximum
in the antiaging regime (see Fig.\ref{fig2}). In the limit
$\beta\rightarrow 0$ or $\beta\rightarrow \infty$, we obtain
$G\left( {\beta ;{\nu _0},{\nu _\infty }} \right)\rightarrow1/2$
indicating the marginal role of the non-Markovian dynamics,
i.e., using Eq.(\ref{eq:G}) $f_c^{NMMV}\to f_c^{SMV}$. Since
the critical noise $f_c^{NMMV}$ depends on $\beta$ only through the
function $G\left( {\beta ;{\nu _0},{\nu _\infty }} \right)$  in the
aging regime, the minimum of $G\left( {\beta ;{\nu _0},{\nu _\infty
}} \right)$ is achieved for $\beta=\beta_m^{aging}$, corresponding
to the maximum of  $f_c^{NMMV}$;  conversely in the antiaging
regime the maximum of  $G\left( {\beta ;{\nu _0},{\nu _\infty }}
\right)$ is achieved for $\beta=\beta_m^{antiaging}$ corresponding
to the minimum of $f_c^{NMMV}$. Let us indicate with $\Delta G_m$
the maximal deviation of the function $G$ from its asymptotic value
$1/2$ achieved in the limit $\beta\rightarrow 0$ and
$\beta\rightarrow \infty$, i.e., \bea \Delta G_m =
\left|\frac{1}{2} - G\left( {\beta _m};{\nu _0}, {\nu _\infty
}\right)\right|. \eea Specifically let us indicate with $\Delta
G_m^{aging}$ the values obtained in the aging regime and with
$\Delta G_m^{antiaging}$ the values obtained in the antiaging
regime.
	
Therefore, the values of $\Delta G_m^{aging}$ and $\Delta
G_m^{antiaging}$ characterize the maximal difference between
$f_c^{NMMV}$ and $f_c^{SMV}$ for the aging regime and antiaging regime,
respectively. We note that since $G$ is independent of the topology
of the underlying network, $\beta_m^{aging}$
($\beta_m^{antiaging}$) at which $f_c^{NMMV}$ is maximized
(minimized), is not affected by  the network topology.
	
While the definition of $G$ given by Eq.(\ref{eq31}) is valid for arbitrary functions $\nu(a)$, the investigation performed in this paragraph is obtained starting from the expression for $\nu(a)$ given by Eq.(\ref{eq3}). However,  here we  conjecture that these results do not qualitatively change for other choices of the function $\nu(a)$ as long as the derivative of  function is monotonic. This is strictly speaking the case in which we can properly use the terms {\em aging} and {\em antiaging} for dynamical evolution as an inflection point in $\nu(a)$ (corresponding to a maximum or minimum in the derivative $d\nu/da$)  will introduce a characteristic scale $a=a^{\star}$ on the dynamics.
		
In order to check this conjecture we have  considered several functions $\nu(a)$ with a monotonic first derivative.
	
In particular, we have considered the linear function
	\begin{eqnarray}
	\nu \left( a \right) = \left\{ \begin{array}{l}
	\beta \left( {\nu _\infty   - \nu _0} \right)a + \nu _0,{\kern 10pt}  a < 1/\beta, \\
	\nu _\infty,{\kern 10pt}  a \ge 1/\beta,
	\end{array} \right. \label{eq41}
	\end{eqnarray}
	the rational function
	\begin{eqnarray}
	\nu \left( a \right) = \frac{{\nu _ \infty a + \nu_0 /\beta }}{{a +
			1/\beta }}, \label{eq42}
	\end{eqnarray}
	and the expression
	\begin{eqnarray}
	\nu \left( a \right) = \left( {\nu_0 - \nu_\infty} \right){\left( {1
			+ a} \right)^{ - \beta }} + \nu_\infty. \label{eq43}
	\end{eqnarray}
including a power-law dependence on the age $a$. Interestingly, the functional  dependences given by Eq.(\ref{eq42}) and Eq.(\ref{eq43})  reproduce non-Markovian dynamics observed in inter-event times and duration of social contact (see for a review Ref.~\cite{Karsai2018}).  We have studied the function $G\left( {\beta ;{\nu_0},{\nu_\infty }}\right)$ for all
these kernels, and we have found that qualitatively  the results are
unchanged with respect to the results obtained for the exponential
kernel.

\begin{figure}
\centerline{\includegraphics*[width=1.0\columnwidth]{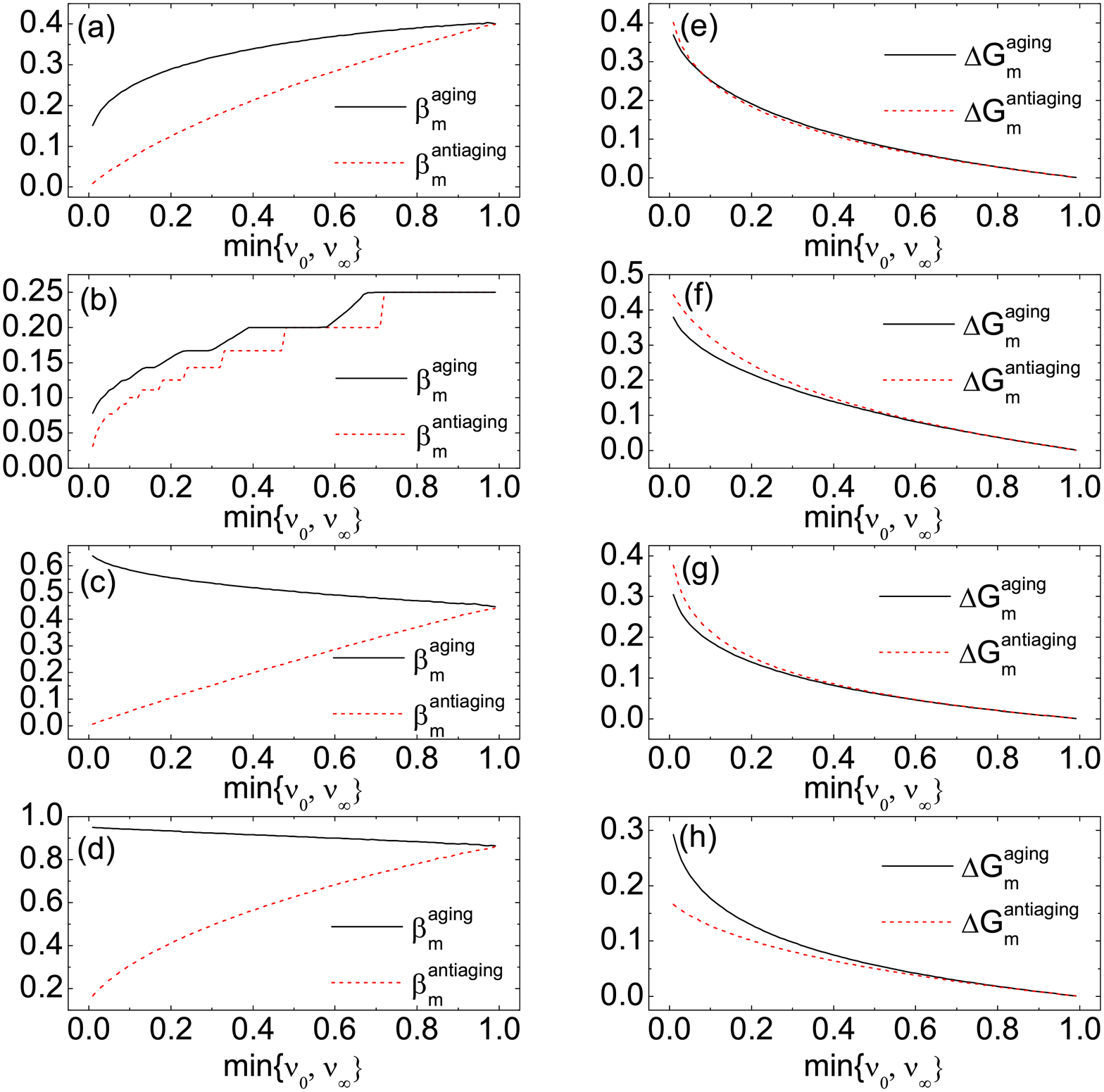}}
\caption{In the left panels, the values of $\beta_m^{aging}$ and  $\beta_m^{antiaging}$ are plotted versus $\min \left\{ {{\nu _0},{\nu _\infty }} \right\}$. In the right panels, the values of  $\Delta G_m^{aging}$ and $\Delta G_m^{antiaging}$ are plotted  as a function of $\min \left\{ {{\nu _0},{\nu _\infty }} \right\}$. From the top to bottom, we consider four functional forms for $\nu(a)$ are considered: exponential (Eq.\ref{eq3}), linear (Eq.\ref{eq41}), rational (Eq.\ref{eq42}), and power-law (Eq.\ref{eq43}) kernels. All curves are obtained for $\max \left\{ {{\nu _0},{\nu _\infty }} \right\}= 1$.
\label{fig3}}
\end{figure}

In Fig.~\ref{fig3} we show the dependence of
$\beta_m^{aging}$, $\beta_m^{antiaging}$ and of $\Delta
G_m^{aging}$ and  $\Delta G_m^{antiaging}$ as a function of $\min
\left\{ {{\nu _0},{\nu _\infty }} \right\}$ for the four types of kernel with a fixed value of $\max
\left\{ {{\nu _0},{\nu _\infty }} \right\}=1$. We observe that while $\Delta
G_m^{aging}$, $\Delta G_m^{antiaging}$ and $\beta_m^{antiaging}$ show the same monotonic trend for all the kernels, $\beta_m^{aging}$ displays a different trend depending on the considered kernels. Indeed while for the linear and exponential kernels both $\beta_m^{aging}$, increase with $\min
\left\{ {{\nu _0},{\nu _\infty }} \right\}$, for the rational and power-law kernels it decreases with $\min
\left\{ {{\nu _0},{\nu _\infty }} \right\}$.

\begin{figure*}
\centerline{\includegraphics*[width=2\columnwidth]{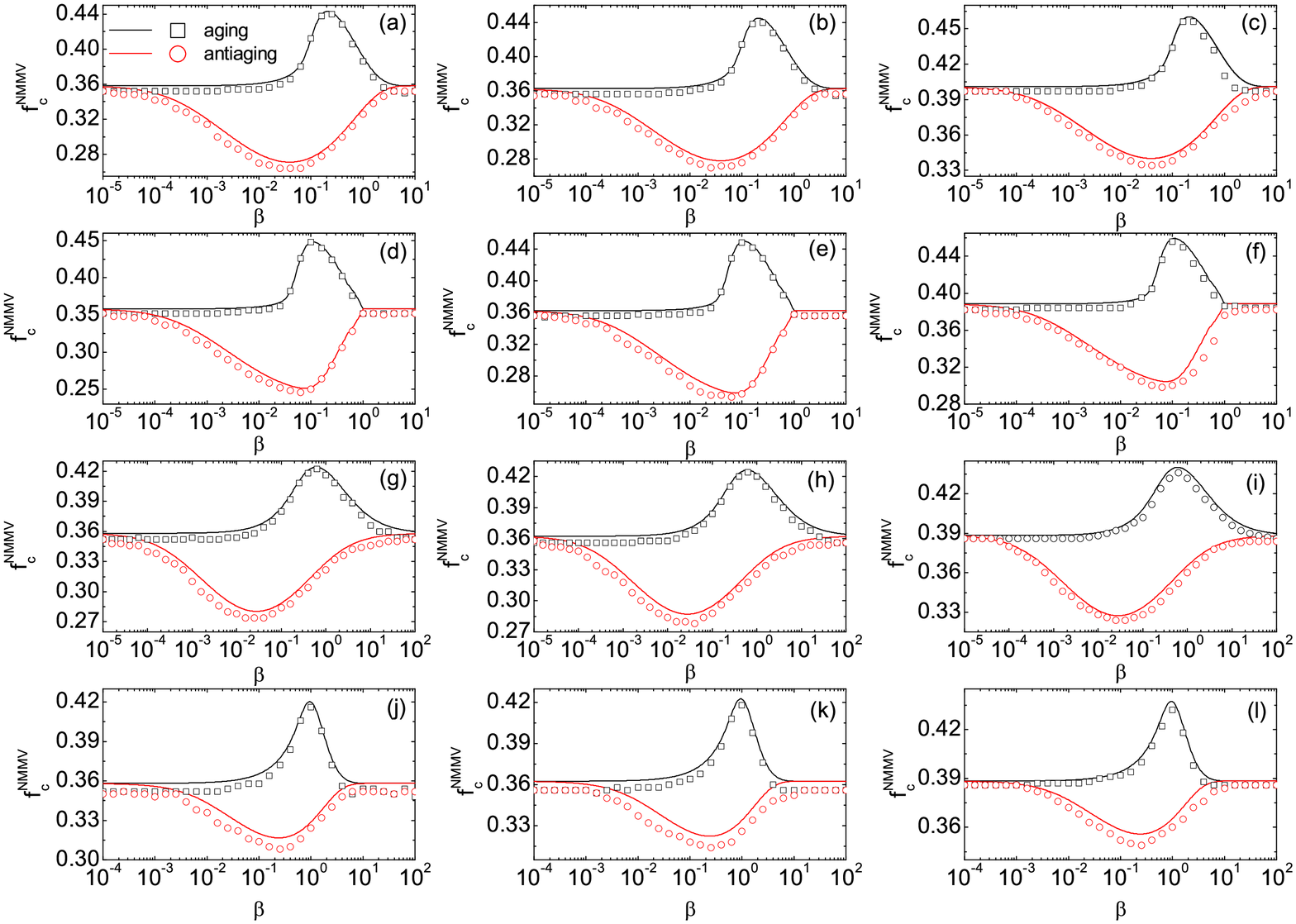}}
\caption{The critical noise $f_c^{NMMV}$ is plotted as a function of $\beta$ in the aging and
antiaging regimes for three different networks (from left to right: the regular-random networks (RR) with degree of each node $\left\langle k \right\rangle = 20$, the Erd\"os-R\'enyi (ER) random networks with average degree $\left\langle k \right\rangle = 20$, and scale-free networks with degree distribution exponent $\gamma=3$ and minimal degree $k_{min}=10$) and four types of $\nu(a)$ (from top to bottom: exponential (Eq.\ref{eq3}), linear (Eq.\ref{eq41}), rational (Eq.\ref{eq42}), and power-law (Eq.\ref{eq43}) kernels). Symbols and lines show the
simulation and theoretical results, respectively. All curves are obtained by setting  $\max \left\{ {{\nu _0},{\nu _\infty }} \right\} = 1$
and $\min \left\{ {{\nu _0},{\nu _\infty }} \right\} = 0.05$. We have
also performed simulations with some other values of  $\min \left\{
{{\nu _0},{\nu _\infty }} \right\}$, and found that the
non-monotonic  behavior of $f_c^{NMMV}$ is qualitatively the same.}
\label{fig4}
\end{figure*}
	
\section{Comparison with numerical results}\label{sec5}
In this section we compare the results obtained analytically using
the heterogeneous mean-field approximation with extensive numerical
results on different network topologies.
	
We have considered three different random networks generated using
the configuration model \cite{PRE.64.026118}:
\begin{itemize}
\item[(a)]
regular random networks
(RR) with degree distribution $P\left( k \right) = \delta \left( {k - \left\langle k \right\rangle } \right)$;
\item[(b)]Erd\"os-R\'enyi networks (ER) with degree distribution
$P\left( k \right) = {{{e^{ - \left\langle
k \right\rangle }}{{\left\langle k \right\rangle }^k}}
\mathord{\left/ {\vphantom {{{e^{ - \left\langle k \right\rangle
}}{{\left\langle k \right\rangle }^k}} {k!}}} \right.
\kern-\nulldelimiterspace} {k!}}$;
\item[(c)] scale-free networks (SF) with degree distribution $P\left( k \right)\sim k^{ -\gamma}$.
\end{itemize}
In order to numerically determine the critical noise $f_c^{NMMV}$,
we calculated the Binder's fourth-order cumulant $U$
\cite{Binder.RPP.1997}, defined as \bea U = 1
-\frac{1}{3}\frac{\left[ {\overline {m^4} } \right]}{\left[
{\overline{m^2} } \right]^2}, \eea where $m = \sum_i^N\sigma_i/N$ is
the average magnetization per node, $\overline{\cdot}$ denotes the
time averages taken in the stationary regime, and $\left[ \cdot
\right]$ indicates the averages over different network
configurations. The critical noise $f_c^{NMMV}$ is obtained by
detecting the point $f=f_c^{NMMV}$  where the curves $U =U(f)$
obtained for different network sizes $N$, intercept each other. In
Fig.\ref{fig4}, we show $f_c^{NMMV}$ as a function of $\beta$ for
the aging and antiaging regime for the three considered network
models and for the four types of considered  kernels without characteristic scale, finding very good agreement with the mean-field theoretical predictions despite these latter neglect the correlations present in the NMMV.
	
As predicted by the mean-field theory, for all the considered choices of $\nu(a)$ without inflection point, the critical noise
$f_c^{NMMV}$ shows a non-monotonic dependence on $\beta$ in both
regimes. In the aging regime, there exists an optimal value of
$\beta$ in which $f_c^{NMMV}$ is maximized, in the antiaging regime
instead $f_c^{NMMV}$ displays a minimum as a function of $\beta$.
The optimal $\beta$ for the two regimes are independent of the
network degree distribution as predicted by the heterogeneous
mean-field solution.
	
However, this scenario can change if the function $\nu(a)$ describes a dynamics with a characteristic scale. This is not typically the scenario considered in physical works investigating the slow down of the dynamics due to aging, but it is actually a very valuable choice in the present context of social opinion dynamics.
To investigate this case here we focus on the  class of logistic functions $\nu(a)$ given by
\bea
\nu(a)=\frac{\nu_0-\nu_{\infty}}{1+e^{\beta (a-a^{\star})}}+\nu_{\infty},
\label{nu:logistic}
\eea
with both $a^{\star}$ and $\beta$ being non-negative.
This logistic function is a monotonic function of $a$ and for large values of $\beta$ approaches a step function at $a=a^{\star}$.  Most notably this choice of functional for, for $\nu(a)$ introduces  a characteristic scale $a=a^{\star}$ for age at which the change of opinion occurs.
		
\begin{figure}
\centerline{\includegraphics*[width=1.0\columnwidth]{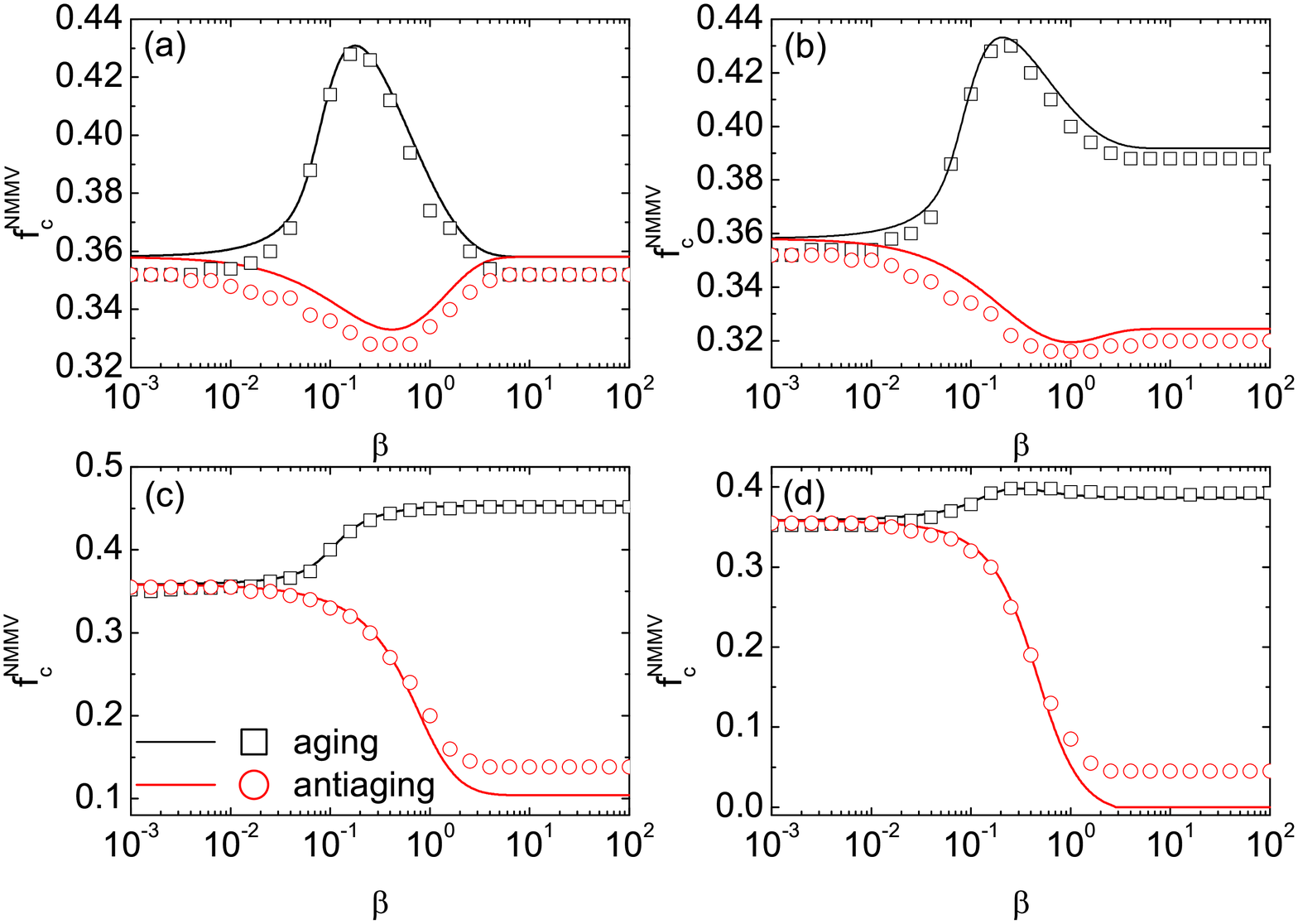}}
\caption{The critical noise $f_c^{NMMV}$ is plotted as a function of $\beta$ in the aging and antiaging regime for four different values of $a^{\star}$ when $\nu(a)$ takes the logistic form given by Eq.(\ref{nu:logistic}). From (a)-(d), $a^{\star}=0$, 1, 5, and 10, respectively.
We have used the regular random networks with degree of each node  given by $k=20$. Symbols and lines show the
simulation and theoretical results, respectively. All results are obtained by setting  $\max \left\{ {{\nu _0},{\nu _\infty }} \right\} = 1$
and $\min \left\{ {{\nu _0},{\nu _\infty }} \right\} = 0.05$.}
\label{fig:logistic}
\end{figure}

We have simulated the NMMV model with this logistic kernel and compared the theory with the analytical mean-field prediction finding satisfactory agreement between the two (see Fig. \ref{fig:logistic}).
		
Interestingly in this case we observe that only for $a^{\star}=0$ (where there is no effective typical scale in the system) we recover the same qualitative behavior of $f^{NMMV}$ observed in the previous kernels (see Fig. \ref{fig4}). We therefore make the important observation that the introduction of a typical scale $a=a^{\star}$ can significantly alter the phenomenology of the process.

\begin{figure}
\centerline{\includegraphics*[width=1.0\columnwidth]{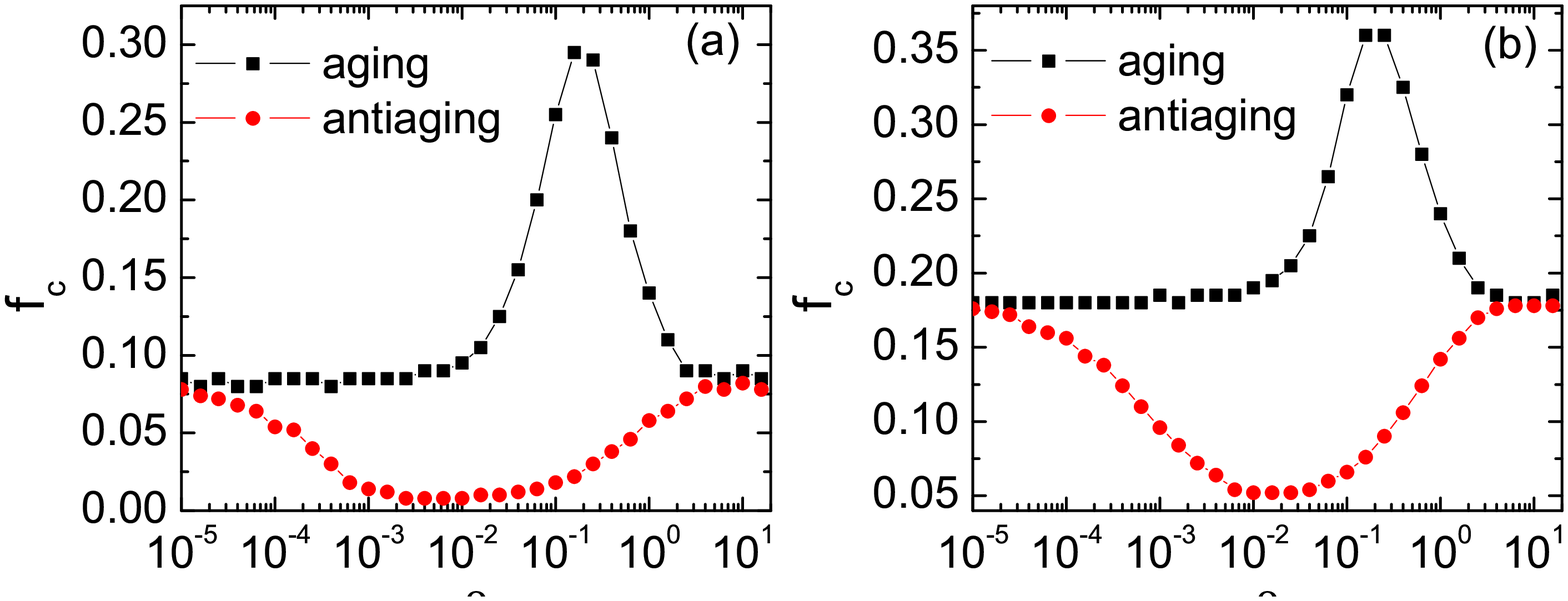}}
\caption{The critical noise $f_c^{NMMV}$ is plotted as a function of $\beta$ in the aging and antiaging regime for a  2d square lattices (panel a) and in 3d cubic lattices (panel b) where we have take the exponential kernel $\nu(a)$ given by Eq. (\ref{eq3}). All results are obtained setting  $\max \left\{ {{\nu _0},{\nu _\infty }} \right\} = 1$
and $\min \left\{ {{\nu _0},{\nu _\infty }} \right\} = 0.05$.
\label{fig5}}
\end{figure}

Finally, we investigated the NMMV model also on two-dimensional and
three-dimensional regular lattices,  which are network topologies
for which the heterogeneous mean-field approximation is not valid.
For these lattices we have exclusively considered the exponential kernel given by Eq. (\ref{eq3}). The results are shown in Fig.\ref{fig5}. For two-dimensional
lattices, the critical noise shows a maximum $f^{NMMV}_c\approx 0.3$
at $\beta_m^{aging} \approx 0.2$ in the aging regime and a minimum
$f^{NMMV}_c\approx 0.008$ at $\beta_m^{antiaging} \approx 0.01$ in
the antiaging regime. For three-dimensional lattices, the critical
noise shows a maximum $f^{NMMV}_c\approx 0.36$ at $\beta_m^{aging}
\approx 0.2$ in the regime regime
and a minimum $f^{NMMV}_c\approx 0.05$ at $\beta_m^{antiaging} \approx 0.01$
in the antiaging regime. In the limits of $\beta\rightarrow 0$ and
$\beta\rightarrow \infty$, the critical noise tend respectively to
$0.075$ and $0.18$ in two-dimensional and three-dimensional
lattices, consistent with the results valid for the SMV model
\cite{PhysRevE.77.051122}. This result shows evidently that also in
situations in which we are far from the conditions necessary for the
application of the heterogeneous approximation we observe a
non-monotonic   dependence of the critical noise $f_c^{NMMV}$ of the
NMMV model on $\beta$ revealing that the observed phenomenology is
universal, i.e., it is independent of the network topology.

\section{Conclusion}\label{sec6}
	
In this work we have introduced the non-Markovian Majority-Vote
(NMMV) model that differs from the standard Majority-Vote (SMV)
model as it includes  memory effects. In fact in the NMMV model the
probability that an agent switches  state  (activation probability)
is not only dependent on the  majority state of its neighbours as
for the SMV model, but it is also  age-dependent, i.e.\
depends on how long a agent has been in the same state (his age $a$) captured by the function $\nu(a)$.
	
We distinguish two regime of the NMMV model: the aging regime in
which the activation probability is a decreasing function of the
agent's age, and the antiaging regime in which the activation
probability is an increasing function of the agent's age. We call
$\beta$ the rate determining the change of the
activation probability with the age of the agent. The NMMV model
displays a phase transition as a function of the noise $f$
determining the probability that an agent switches to the minority
state of its neighbors. For $f<f_c^{NMMV}$ the NMMV model is in an
ordered phase and displays an overall majority state, for $f\geq
f_c^{NMMV}$ the model is in a disordered phase in which half
of the agents are in one state and the other half of the agents are in the other state.
	
By analytically solving the model using the heterogeneous mean-field
approach and by performing extensive numerical simulations, we reveal how the non-Markovian dynamics affects the critical noise $f_c^{NMMV}.$
	
These results indicate that in the aging regime the non-Markovian
dynamics retards the transition, and in the antiaging dynamics it
anticipates the transition. Interestingly the most significant
effect of the non-Markovian dynamics is achieved at a finite and
non-zero value of the rate $\beta$, indicating that the
aging/antiaging dynamics needs to have a characteristic time-scale
that is neither  too fast or too slow.
	
Interestingly,  as long as the non-Markovian kernel $\nu(a)$  does not have a characteristic scale, the critical noise $f_c^{NMMV}$ in the NMMV model exhibits a non-monotonic dependence on the rate
$\beta$ at which the activation probability changes with age. In particular we found two opposite behaviors in the aging and in the antiaging regimes. In the aging regime, the critical noise $f_c^{NMMV}$ displays a maximum as a function of $\beta$  in the antiaging regime instead  $f_c^{NMMV}$ displays a minimum as a function of $\beta$.

Finally, this work highlights the importance of non-Markovian
dynamics  in determining the phase diagram of the NMMV model and we
hope that it will stimulate interest in further investigations of
the effect of memory and non-Markovian dynamics  in critical
phenomena defined on networks.

\begin{acknowledgments}
We acknowledge the anonymous referee for pointing out the use of the
logistic kernel in Eq.(\ref{nu:logistic}). This work was supported by the National Natural Science Foundation of China (11875069, 11975025, 12011530158, 61973001) and the Royal Society (IEC\textbackslash NSFC \textbackslash 191147)
\end{acknowledgments}

\end{document}